\documentclass[11pt]{article}
\usepackage[utf8]{inputenc}

\usepackage{fullpage}

\usepackage{amsmath}
\usepackage{amssymb}
\usepackage{amsthm}
\usepackage{authblk}

\usepackage{xfrac}
\usepackage{algpseudocode}
\usepackage{algorithm}
\usepackage{tikz}
\usetikzlibrary{positioning}
\usepackage{xcolor}

\usepackage{hyperref}
\usepackage{cleveref}%
\usepackage{lipsum}%

\title{Evaluating the Claims of ``$\sat$ Requires Exhaustive Search''\thanks{Supported in part by NSF grant 
		CCF-2006496.}}

\author{Michael C. Chavrimootoo}
\author{Yumeng He}
\author{Matan Kotler-Berkowitz}
\author{Harry Liuson}
\author{Zeyu~Nie}
\affil{Department of Computer Science\\University of Rochester\\Rochester, NY 14627, USA}

\newcommand{\p}{\ensuremath{{\rm P}}}
\newcommand{\np}{\ensuremath{{\rm NP}}}

\newcommand{\sat}{\ensuremath{{\rm SAT}}}
\newcommand{\csp}{\ensuremath{{\rm CSP}}}

\newtheorem{theorem}{Theorem}
\newtheorem{lemma}{Lemma}

\newtheorem{definition}{Definition}
\date{December 4, 2023}

\begin{document}\sloppy

\maketitle

\begin{abstract}
In this paper, we take a closer look at the claims made by Xu and Zhou in their paper ``SAT Requires Exhaustive Search''~\cite{xu-zho:t:sat-exp}, which claims to provide a lower bound on the complexity of the so-called Model~RB\@. Xu and Zhou conclude that their result implies a separation between P and NP, since the lower bound purportedly proves that the Strong Exponential Time Hypothesis (SETH) is true. In examining Xu and Zhou's arguments, we find a flaw in their main theorems. The authors assume that an algorithm for Model~RB must have a certain structure that can leverage downward self-reducibility, and argue that such an algorithm cannot run in polynomial time. We argue that this structure is not guaranteed to exist and thus their paper neither proves SETH to be true nor proves $\p \neq \np$.
\end{abstract}

\section{Introduction}

This critique provides an analysis of Xu and Zhou's ``SAT Requires Exhaustive Search''~\cite{xu-zho:t:sat-exp}. Xu and Zhou attempt to prove that a particular form of constraint satisfaction problems, which they refer to as Model~RB, cannot be solved without exhaustive search. While that claim is interesting in its own right, we give it special consideration as it further claims to prove that $\p \neq \np$\@. Indeed, the paper claims that for each positive real-valued constant $c < 1$, SAT cannot be solved in $O(2^{cn})$ time, which is a stronger claim than separating $\p$ from $\np$: it is equivalent to the Strong Exponential Time Hypothesis (SETH)\@! If SETH is true, many known problems are only solvable through a brute force algorithm, including generalized CNF satisfiability and the maximum cut graph problem~\cite{mih-wil-rya:c:possibility-faster-sat}.
We thus explore the arguments of Xu and Zhou~\cite{xu-zho:t:sat-exp} and identify that they incorrectly rely on several unproven assumptions about the nature of Model~RB and about the potential algorithms for solving it.

In Section~\ref{s:prelims}, we outline Model~RB and the key theorems in Xu and Zhou's paper and discuss how they use Model~RB to prove that certain constraint satisfaction problems cannot be solved without exhaustive search. Next, in Section~\ref{s:main} we analyze the arguments that Xu and Zhou make about how instances of Model~RB must be solved, and find two issues. First, they assume without proof that instances of Model~RB are self-reducible; we demonstrate that with high probability this assumption is incorrect. Second, they rely on several unproven assumptions that limit the potential algorithms for solving instances of Model~RB\@. Finally, in Section~\ref{s:conclusion} we conclude that the flaws identified in Section 3 mean that Xu and Zhou's claim that $\p \neq \np$ does not follow from their arguments.

\section{Preliminaries}\label{s:prelims}

As is standard we use, $\ln(\cdot)$ as a shorthand for $\log_e(\cdot)$, i.e., the natural log.

\subsection{CSP and Model~RB}
    A \textit{Constraint Satisfaction Problem (CSP)} can be defined as a triple $(X,D,C)$, where
    \begin{itemize}
        \item $X=\{x_1,x_2,\ldots,x_n\}$ is a set of variables,
        \item $D=\{D_1, D_2,\ldots,D_n\}$ is a set of domains of values such that each $x_i$ takes on values from $D_i$, where the domain $D_i$ can be a set of objects of any type, and
        \item $C=\{C_1, C_2,\ldots, C_m\}$ is a set of constraints, where each constraint $C_i=(X_i, R_i)$, with $X_i=\{x_{i_1},x_{i_2},\ldots,x_{i_k}\}\subseteq X$ is a subset of $k$ variables and $R_i\subseteq D_{x_{i_1}} \times \cdots\times D_{x_{i_k}}$ is the permitted set of tuples of values from the corresponding $k$ domains. So the relation $R_i$ limits the combinations of values that the $k$ variables in $X_i$ can take to only those that are permitted by the constraint.
    \end{itemize}

    An assignment $\sigma\in D_1\times\cdots\times D_n$ is said to satisfy a constraint $C_j=(X_j,R_j)$ if the values assigned to $X_j$ satisfy the relation $R_j$. An assignment $\sigma$ is said to satisfy the CSP if it satisfies all the constraints.
    
    \textit{Model~RB} is a ``random CSP model,'' i.e., it generates random instances of CSP, and was proposed by Xu and Li~\cite{xu-li:j:phase-csp}. Adhering to the parameters defined above for a general CSP, a random CSP instance $I$ of Model~RB has the additional properties below~\cite{xu-zho:t:sat-exp}:
    \begin{itemize}
        \item $(\forall i\in\{1,\ldots,n\})[
        \|D_i\|=d=n^\alpha]$, where $\alpha>0$ is a constant,
        \item $m=rn\ln d$, where $r>0$ is a constant,
        \item $(\forall i\in\{1,\ldots,n\})[\|X_i\|=k]$, where $k\ge 2$ is a constant, and the $k$ distinct variables are chosen uniformly at random from $X$, and
        \item $(\forall i\in\{1,\ldots,n\})[\|R_i\|=(1-p)d^k]$, where $0<p<1$ is a constant, and each tuple of values is selected uniformly at random from $D_{x_{i_1}} \times \cdots\times D_{x_{i_k}}$.
    \end{itemize}

    From the properties of CSP instances generated by Model~RB, we can see that the domain size $d$ grows with the number of variables $n$, therefore we have the property that $O(1/d)=o(1)$ which is used in Theorem~2.5 of Xu and Zhou's paper~\cite{xu-zho:t:sat-exp}.

    Furthermore, they define a \textit{symmetry requirement} for the permitted set $R_i$ of each constraint $C_i$ by defining how those permitted sets are generated. More specifically, they begin with a ``symmetry set'' $R$ that contains $(1-p)d^k$ tuples of values, and from $R$ generate each $R_i$ by applying random permutations to the domains of $k-1$ of the variables in $X_i$.

    They also define the \textit{symmetry mapping} of a constraint $C_i=(X_i,R_i)$ in Definition~\ref{d:sym-mapping} (see the next definition), which changes the constraint's permitted set $R_i$ slightly by interchanging two values of a variable $x_j\in X_i$.
    
    \begin{definition}[{\protect\cite[Definition~2.2]{xu-zho:t:sat-exp}}] \label{d:sym-mapping}
    Consider a random instance $I$ of Model~RB with $k = 2$. Assume that $C=(X, R)$ is a constraint of $I$ and $X=\{x_1,x_2\}$, then a \textit{symmetry mapping} of $C$ is to change $R$ by choosing $u_1,u_2\in D_1$ such that $u_1\ne u_2$, and choosing $v_1,v_2\in D_2$ such that $v_1\ne v_2$, where $(u_1,v_1),(u_2,v_2)\in R$ and $(u_1,v_2),(u_2,v_1)\notin R$, and then exchange $u_1$ with $u_2$.
    \end{definition}

\subsection{Important Arguments in the Paper of Xu and Zhou}
 Xu and Zhou~\cite{xu-zho:t:sat-exp} build up their arguments mainly based on lemmas in their Section~2 and arrive at Theorem~\ref{t:RB-not-in-P} as the main conclusion. We list some important arguments in their paper that help to understand their reasoning. 

The following lemma bounds the probability of an instance of Model~RB to be satisfiable.

\begin{lemma}[{\protect\cite[Lemma~2.2]{xu-zho:t:sat-exp}}] \label{l:bound-satisfiable}
    Let $I$ be a random CSP instance of Model~RB\@. Then
    $$ \frac{1}{3} \leq \Pr(I \text{ is satisfiable}) \leq \frac{1}{2}$$
\end{lemma}

In this paper, we use this lemma in the next section to find an upper bound for the number of solutions of subproblems of Model~RB\@.

In their Theorem~2.5, their usage of the notion of ``fixed point" and the meaning of ``symmetry mapping of changing satisfiability" is not clear, so what they tried to show can be put in a more precise way. To avoid ambiguity, we import their original theorem as the following one, with our interpretation below.
\begin{theorem}[{\protect\cite[Theorem~2.5]{xu-zho:t:sat-exp}}] \label{t:symmetry-mapping}
    There exists an infinite set of satisfiable and unsatisfiable instances of Model~RB
    such that this set is a fixed point under the symmetry mapping of changing satisfiability.
\end{theorem}

According to their proof, we interpret this theorem as stating the existence of an infinite set $S$ of instances of Model~RB, such that for each instance $I$ in $S$, (1)~if $I$ is satisfiable (i.e. has at least one solution), then there exists a symmetry mapping on $I$ that maps $I$ to an unsatisfiable instance (i.e. has no solution) in $S$, and (2)~if $I$ is unsatisfiable, then there exists a symmetry mapping on $I$ that maps $I$ to a satisfiable instance in $S$.

Their proof of this theorem relies on the symmetry requirement of Model~RB and uses symmetry mapping to eliminate (or add) exactly one solution from instances of Model~RB that have exactly one (or zero) solution(s) with high probability. Note that since their proof relies on the probability bounds on instances of Model~RB provided by some former lemmas, this theorem cannot be generalized to CSPs that are not instances of Model~RB\@.

Based on divide-and-conquer algorithms for solving CSP, they have a lemma stating the following.
\begin{lemma}[{\protect\cite[Lemma~3.1]{xu-zho:t:sat-exp}}] \label{l:CSP-time-bound}
    If a CSP problem with $n$ variables and domain size $d$ can be solved in time \(T(n) = O(d^{cn})\) time (where $0 < c < 1$), then at most \(O(d^c)\) subproblems with \( n - 1\) variables are needed to solve the original problem.
\end{lemma}

As their main conclusion, they state the following theorem.
\begin{theorem}[{\protect\cite[Theorem~3.2]{xu-zho:t:sat-exp}}] \label{t:RB-not-in-P}
    Model~RB cannot be solved in time $O(d^{cn})$ time for every constant $0 < c < 1$.
\end{theorem}
Their purported proof is based on a false assumption that only divide-and-conquer algorithms can solve CSP\@. In this paper, we will focus on evaluating their claimed proof of that theorem.

\section{Analysis of the Arguments\label{s:main}}

In this section, we identify several issues in the arguments made by Xu and Zhou~\cite{xu-zho:t:sat-exp} and argue that they fail to establish the results in that paper's Section~3.

\subsection{Analysis of Subproblem Difficulty}
The difficulty of Model~RB is argued in Theorem~\ref{t:RB-not-in-P} using the argument
that if not all possible variable assignments are attempted, it is with high probability possible to use the symmetry mapping to invert the satisfiability of the problem without changing any results of the investigated assignments, thus producing a contradiction. Thus, in order to solve an instance of Model~RB, we must solve $d$ subproblems with $n-1$ variables. However, the authors seem to assume without proof that each of these subproblems is also an instance of Model~RB, which also produces $d$ subproblems, and so on, requiring an exhaustive search on the order of $O(d^n)$. 

Assigning an arbitrary value produces a subproblem with $n-1$ variables, each of which has $d$ possible values. We have on average $$\frac{\binom{n-1}{k-1}}{\binom{n}{k}} rn \ln d= rk \ln d$$ constraints with $k-1$ variables and $r(n-k)\ln d$ constraints with $k$ variables.

Clearly, the probability of satisfaction for constraints with $k$ variables remains $1-p$. For the remaining constraints, we have reduced the relevant possibility space to size $d^{k-1}$ as well as reduced the number of permitted tuples to $(1-p)d^{k-1}$, which gives us again a probability of satisfaction of $1-p$.

Then, for some given variable assignment of the subproblem, the probability of satisfiability remains $(1-p)^{rn \ln d}$. By Lemma~\ref{l:bound-satisfiable}, we can upper bound the probability of satisfiability $X$ for the subproblem by the expected number of solutions, thus
\[P(X > 0) \leq d^{n-1} (1-p)^{rn \ln d} = \frac{1}{2d}.\]

Therefore, these subproblems lack the crucial feature of Model~RB (with constants carefully selected), which is that it toes the line very narrowly between satisfiability and unsatisfiability. In particular, they become increasingly unlikely to be satisfiable as $d$ grows as a polynomial function of $n$. When subproblems are very likely to be unsatisfiable, we may be able to prune large swathes of the search space away, and as such it is not possible to conclude that exhaustive search is required.

\subsection{Analysis of Subproblem Requirement and Generation}
Lemma~\ref{l:CSP-time-bound} arrives at its conclusion by reformulating the time complexity of the original problem, \(T(n) = O(d^{cn})\), into the complexity \[T(n) = O(d^c) \times T(n - 1).\] This reformulated time complexity is equivalent to multiplying the number of subproblems \(O(d^c)\) by the time complexity of each subproblem  \(T(n - 1)\)\@. This suggests that if a subproblem-based algorithm were used to solve the original CSP, at most \(O(d^c)\) subproblems would be required.

However, Theorem~\ref{t:RB-not-in-P} uses Lemma~\ref{l:CSP-time-bound} to assume that the \textit{only} method of solving the original problem is by breaking it into subproblems. This assumption does not have a proof and may be incorrect. There may exist another algorithm to solve the original CSP that does not involve dividing-and-conquering the CSP into subproblems;\footnote{Of course, we cannot prove in this paper whether such an algorithm actually exists, as this would resolve the P versus NP problem.} the fact that the time complexity of the original problem can be written as a multiple of the time complexity of its subproblem does not mean that the only algorithm to solve the original problem is indeed subproblem-based.

Furthermore, let us assume that divide-and-conquer through subproblems is the only correct approach to solving the original CSP problem. Theorem~\ref{t:RB-not-in-P} still assumes, without a proof, that it must use a particular version of divide-and-conquer. Specifically, it assumes that it must generate the subproblems by choosing an arbitrary variable, then for each of those subproblems replacing that variable with a different constant in its domain. This assumption again does not have a proof, and other subproblem generation strategies could yield more efficient algorithms. For instance, an alternative approach is to generate subproblems by choosing a different variable for each subproblem. While this alternative strategy would also almost certainly not enable solving the original problem in polynomial time, the fact that other subproblem generation strategies exist demonstrates that this assumption is incorrect.

Theorem~\ref{t:RB-not-in-P} relies on both of these assumptions to demonstrate that Model~RB cannot be solved in \(O(d^{cn})\) time. In particular, limiting the set of possible algorithms allows Theorem~\ref{t:RB-not-in-P} to claim that a Model~RB instance may be symmetry-mapped to change its satisfiability without changing the subproblems. Therefore, Theorem~\ref{t:RB-not-in-P} argues that a subproblem-based algorithm that searches a polynomial number of subproblems cannot solve the original problem, and so the problem cannot be solved in polynomial time. However, this claim only applies to this particular type of algorithm. If the possible CSP algorithms are not limited to this particular method of divide-and-conquer, then the central claim of Theorem~\ref{t:RB-not-in-P} does not hold. 

\subsection{Connections to Complexity and Analysis of Complexity Results}

We mentioned in our introduction that Xu and Zhou~\cite{xu-zho:t:sat-exp} claim that their main result (i.e., Theorem~\ref{t:RB-not-in-P}) implies that for each positive real-valued constant $c < 1$, $\sat$ cannot be solved in $O(2^{cn})$, which is equivalent to the Strong Exponential Time Hypothesis (see our introduction for a list of references and mentions of its importance). The purported proof of that corollary (which itself is described in the paragraph before the corollary) discusses a polynomial-time encoding by Walsh~\cite{wal:c:sat-csp} that encodes instances of $\csp$ (Constraint Satisfaction Problems) into instances of $\sat$\@. However, Walsh's encoding is a mapping between two decision problems, whereas Model~RB defines a distributional problem, i.e., a pair $(L, \mu)$, where $L$ is a language and $\mu$ is a distribution (see~\cite[Chapter~18]{aro-bar:b:complexity} for an introduction to distributional problems).  
(We note in passing that the decision problem that underlies the distributional problem defined by Model~RB does seem to be CSP\@.)
Xu and Zhou's paper~\cite{xu-zho:t:sat-exp} is trying to use Walsh's encoding to create a deterministic reduction from a distributional problem to a decision problem. While we are not claiming that this reduction does not exist, we do observe that the direct application of Walsh's encoding does not establish a meaningful relationship between the distributional problem defined by Model~RB and $\sat$\@.

\section{Conclusion}\label{s:conclusion}
While we find no fault in Section 2 of Xu and Zhou~\cite{xu-zho:t:sat-exp}, from our analysis we find that the conclusion $\p \neq \np$ does not follow readily from the results of Section 2. First, as mentioned above, their claim contradicts known results that it is possible to solve 3-SAT problems without exhaustive search, but still in exponential time. Second, they attempt to show that when using a divide-and-conquer strategy, Model~RB requires $O(d^{n})$ time. Even supposing this were true, it does not preclude the existence of alternative algorithms that might solve instances of Model~RB in 
time $O(d^{cn})$ for some positive real-valued constant $c < 1$. And finally, even using a divide-and-conquer strategy, it is not possible to conclude that exhaustive search is required. It seems to be true indeed that Model~RB must be reduced to a number of subproblems $d$, which is polynomial in $n$. This implies exhaustive search at the top level of the search tree. However, these subproblems are not instances of Model~RB, and instead are more general instances of CSP\@. As such, the entire purported proof strategy of attempting to establish a recurrence relation on the size of the problem is not valid.

\paragraph{Acknowledgements}
We would like to thank
Lane A. Hemaspaandra,
Tran Duy Anh Le, and
Eliot J. Smith
for their helpful comments on prior drafts.
The authors are responsible for any remaining errors.

\bibliographystyle{alpha}
\bibliography{gry-reu,local_refs}

\begin{thebibliography}{Wal00}

\bibitem[AB09]{aro-bar:b:complexity}
S.~Arora and B.~Barak.
\newblock {\em Complexity Theory: A Modern Approach}.
\newblock Cambridge University Press, 2009.

\bibitem[PW10]{mih-wil-rya:c:possibility-faster-sat}
Mihai P\u{a}tra\c{s}cu and Ryan Williams.
\newblock On the possibility of faster {SAT} algorithms.
\newblock In {\em Proceedings of the Twenty-First Annual ACM--SIAM Symposium on Discrete Algorithms}, pages 1065--1075, USA, 2010. Society for Industrial and Applied Mathematics.

\bibitem[Wal00]{wal:c:sat-csp}
T.~Walsh.
\newblock {SAT} v {CSP}.
\newblock In {\em Proceedings of the 6th International Conference of Principles and Practice of Constraint Programming}, pages 441--456. Springer-Verlag {Lecture Notes in Computer Science \#1894}, September 2000.

\bibitem[XL00]{xu-li:j:phase-csp}
K.~Xu and W.~Li.
\newblock Exact phase transitions in random constraint satisfaction problems.
\newblock {\em Journal of Artificial Intelligence Research}, 53:93--103, April 2000.

\bibitem[XZ23]{xu-zho:t:sat-exp}
K.~Xu and G.~Zhou.
\newblock {SAT} requires exhaustive search.
\newblock Technical Report arXiv:2302.09512~[cs.CC], Computing Research Repository, \mbox{arXiv.org/corr/}, February 2023.
\newblock Revised September 21, 2023.

\end{thebibliography}

\end{document}